\documentclass[twocolumn,floatfix,tightenlines,preprintnumbers,amsmath,amssymb,superscriptaddress]{revtex4} 
\usepackage{url}
\usepackage{hyperref}
\usepackage{epsfig,colordvi}
\usepackage{graphicx}
\usepackage{multirow}
\usepackage{booktabs, ctable}
\usepackage{xspace}
\usepackage{times}
\usepackage{bm}
\usepackage{soul}
\usepackage{ulem}
\usepackage{xcolor}
\usepackage{color}
\usepackage{natbib}
\usepackage{float}
\usepackage{footnote}

\begin{document}

\title{Relativistic corrections to hadron-hadron scattering phase shift and correlation function}

\author{Zeyu Zeng}
\affiliation{Department of Physics, University of Illinois Urbana-Champaign, Urbana, 61801 Illinois, USA}
\author{Baoyi Chen}
\affiliation{Department of Physics, Tianjin University, Tianjin 300354, China}
\author{Jiaxing Zhao}
\affiliation{Helmholtz Research Academy Hesse for FAIR (HFHF), GSI Helmholtz Center for Heavy Ion Physics, Campus Frankfurt, 60438 Frankfurt, Germany}
\affiliation{Institut f\"ur Theoretische Physik, Johann Wolfgang Goethe-Universit\"at,Max-von-Laue-Straße 1, D-60438 Frankfurt am Main, Germany}

\date{\today}

\begin{abstract}
Femtoscopy offers a sensitive probe of hadron emission sources and hadronic interactions. In this study, we examine relativistic corrections to scattering phase shifts and correlation functions using the two-body Dirac equation framework. We analyze the impact of the Darwin term and spin-dependent potentials, showing that these relativistic effects, especially spin-related interactions, significantly enhance the proton-proton correlation function. Our findings emphasize the necessity of including relativistic corrections for precise femtoscopic analyses.
\end{abstract}


 \maketitle
\section{Introduction}
\label{sec.intro}
 
Femtoscopy is a powerful experimental technique for investigating the space-time evolution of the particle-emitting source created in high-energy collisions, as well as for exploring hadron-hadron interactions. This approach has been extensively applied on both the experimental side~\cite{STAR:2005rpl,STAR:2018uho,STAR:2014dcy,STAR:2015kha,ALICE:2018ysd,ALICE:2021cpv,ALICE:2019buq,ALICE:2019hdt,ALICE:2022uso,ALICE:2011kmy} and theoretical side~\cite{Morita:2014kza,Haidenbauer:2018jvl,Kamiya:2019uiw,Kamiya:2022thy,Liu:2022nec,Yan:2024aap,Liu:2024nac,Wang:2024bpl,Xu:2024dnd}. The method is inspired by the Hanbury Brown and Twiss (HBT) effect~\cite{HBTc,Pratt:1984su}, originally developed in astrophysics, and has been widely adopted to probe the space-time structure of the particle-emitting source in nuclear and particle collisions~\cite{Lisa:2005dd,NA49:2007fqa,Li:2008qm,Wiedemann:1996ig,Kisiel:2006is,Xu:2024dnd}.

The key experimental observable in femtoscopy is the two body or three-body momentum correlation function, as shown in Eq.~\eqref{eq.correlation}, where $P({\bf p}_i)$ and $P({\bf p}_1,{\bf p}_2)$ are momentum distribution of single hadron and two hadrons, respectively. Thanks to the advancement of experimental techniques and detector capabilities, numerous two-hadron correlations have been observed in both proton-proton and relativistic heavy-ion collisions at the Relativistic Heavy Ion Collider (RHIC)~\cite{STAR:2005rpl,STAR:2018uho,STAR:2014dcy,STAR:2015kha} and the Large Hadron Collider (LHC)~\cite{ALICE:2018ysd,ALICE:2021cpv,ALICE:2019buq,ALICE:2019hdt,ALICE:2022uso,ALICE:2011kmy}. Within the framework of femtoscopy, these experimentally measured correlation functions can be theoretically described by convoluting the source emission function $S({\bm r})$ with the two-body scattering wavefunction $\psi({\bm r},k)$ via~\cite{Koonin:1977fh,Pratt:1990zq}, 
\begin{eqnarray}
\label{eq.correlation}
C(k)&\equiv&{P({\bf p}_1,{\bf p}_2)\over P({\bf p}_1)P({\bf p}_2)},\nonumber\\
&=&\int S({\bm r})|\psi({\bm r},k)|^2d{\bm r},
\end{eqnarray}
where $k=|{\bf p}^*_1-{\bf p}^*_2|/2$ is the relative momentum in the center-of-mass (CoM) frame of the pair and ${\bf p}^*_i$ is the momentum in the CoM frame. ${\bm r}$ denotes the relative distance between the two particles in the CoM frame. The two-body scattering wave function $\psi({\bf r},k)$ can be obtained by solving the Schr\"odinger equation in the coordinate space or equivalently through the Lippmann-Schwinger equation in momentum space. In this analysis, the commonly adopted assumption of a momentum-independent emission source is applied.
Assuming that the interaction potential is short-ranged, the low-energy scattering properties — and consequently the correlation function — can be effectively characterized by two physical parameters: the scattering length, $a_0$, and the effective range, $r_{\rm eff}$. Furthermore, under the additional assumption that the emission source follows a Gaussian spatial distribution, the correlation function admits an analytical expression, which was first obtained by Lednick\'y and Lyuboshits model~\cite{Lednicky:1981su,Ohnishi:2016elb,ALICE:2018ysd} and has been widely adopted for interpreting experimental correlation data. For realistic interaction potentials and emission sources beyond the Gaussian approximation, the correlation function can be computed numerically using dedicated tools such as the Correlation Afterburner (CRAB)~\cite{crabcite} and the Correlation Analysis Tool using the Schr\"odinger equation (CATS)~\cite{Mihaylov:2018rva,Fabbietti:2020bfg}.

So far, most of the experimentally measured correlation functions have been obtained for pairs of light hadrons, such as $pp$, $pK$, $KK$, and so on. In these systems, the hadron masses are either small or comparable to their momenta. Moreover, the role of spin-dependent interactions remains not fully understood, adding further complexity to the interpretation of these correlations. Under such conditions, relativistic effects can become significant and must be taken into account when calculating the scattering wave functions. 
For bound-state problems, a self-consistent way to incorporate relativistic effects is provided by the covariant Dirac equation, which is particularly suitable for fermionic systems. More generally, the covariant wave equation developed by Sazdjian~\cite{Sazdjian:1986aw,Sazdjian:1988be}, provides a systematic approach for obtaining relativistic bound-state wave functions. The two-body and three-body Dirac equation have been used to investigate the properties of mesons~\cite{Crater:2002fq,Crater:2008rt,Shi:2013rga} and baryons~\cite{Whitney:2011aa,Shi:2019tji}. In contrast, for scattering states, the magnitude of relativistic corrections to experimentally observable quantities and the optimal method for incorporating such effects remain open questions. Previous studies based on relativistic constraint dynamics have extended to phase shift analyses of nucleon-nucleon~\cite{Liu:2002cn} and meson-meson~\cite{Crater:2004xe} scattering. The relativistic correction to the correlation function, obtained by employing a relativistic propagator in momentum space within the Koonin–Pratt formalism, has been investigated in Ref.~\cite{Albaladejo:2024lam}.
In the present work, we propose an alternative and systematic approach in coordinate space to incorporate relativistic corrections to both the scattering phase shifts and the correlation functions.

In this paper, we employ the two-body Dirac equation to investigate two-body scattering and systematically explore the relativistic corrections to both phase shifts and correlation functions, which are experimentally accessible observables.
The paper is organized as follows. The theoretical framework is presented in Section~\ref{sec.frame}, including a brief description to the two-body Dirac equation. Next the calculation on the scattering phase shifts and correlation functions are described. Numerical results and a detailed discussion are provided in Section~\ref{sec.res}. Finally, a summary is given in Section~\ref{sec.sum}.

\section{Theoretical framework}
\label{sec.frame}
In this section, we provide an overview of the theoretical framework adopted to describe the relativistic two-body system. This framework is based on the covariant two-body Dirac equation, which governs the dynamics of spin-1/2 particles interacting via scalar, vector, or other effective potentials (see, for example, Ref.~\cite{10.1063/1.532311}). We outline the essential aspects of the formalism, including the general form of the two-body Dirac equation, the types of interactions considered, and the approach used to obtain both bound and scattering state solutions.

To properly describe scattering processes, we further introduce the methods employed for calculating scattering phase shifts and two-particle correlation functions. These quantities serve as essential observables in experimental studies, providing insight into the interaction dynamics and space-time structure of the particle-emitting source.

\subsection{Two-body Dirac equation}
In this work, for simplicity and relevance to the physical systems under consideration, we restrict our discussion to scalar and vector interactions, which play dominant roles in many hadronic and atomic systems. Under this assumption, the covariant two-body Dirac equation takes the form~\cite{Crater:1983ew,Crater:1987hm,Liu:2002cn}:
\begin{equation}
\begin{aligned}\label{Two-BodyDEq}
    \mathcal{S}_1 \psi &= \gamma_{51}(\gamma_1 \cdot (p_1 - A_1) + m_1 + S_1)\psi = 0\\
    \mathcal{S}_2 \psi &= \gamma_{52}(\gamma_2 \cdot (p_2 - A_2) + m_2 + S_2)\psi = 0
\end{aligned}
\end{equation}
With the 16-component Dirac spinner $\psi \equiv(\psi_1,\psi_2,\psi_3,\psi_4)^T$.
In this equation, $p_1$ and $p_2$ represent the momenta of the two particles, while $A_1$ and $A_2$ are the corresponding vector potentials. The quantities $m_1$ and $m_2$ denote the masses of the two particles, and $S_1$ and $S_2$ represent the scalar potentials acting on the two particles, respectively.
An important feature of this relativistic formalism is that the spin dependence of the interaction emerges naturally from the structure of the relativistic potentials and the mutual compatibility conditions imposed by the two coupled Dirac equations. This stands in contrast to approaches based on semi-relativistic corrections, which typically incorporate spin-dependent terms in an ad hoc manner inspired by non-relativistic approximations or effective field theory. In the two-body Dirac formalism, the relativistic spin structure is consistently embedded from the outset, providing a more fundamental description of the dynamics of interacting fermions.

To simplify the solution of the two-body Dirac equation, H. Crater et al.~\cite{Crater:1983ew,Crater:1987hm,Liu:2002cn} reformulated the equation into a Schr\"odinger-like form by making appropriate combinations of the components of the wave function, such as $\phi_{\pm} = \psi_1 \pm \psi_4$ and $\chi_{\pm} = \psi_2 \pm \psi_3$, followed by appropriate substitutions. This allows the decoupling of the system into four Schr\"odinger-like equations. 
Working in the center-of-mass frame in which $\hat P= (1,0)$ and $\hat r= (0,\hat {\bf r})$ and then further defining four component wave functions $\psi_\pm$, $\eta_\pm$ via $\phi_\pm=\exp(F+K\pmb{\sigma}_1\cdot\hat{\mathbf{r}}\pmb{\sigma}_2\cdot\hat{\mathbf{r}})\psi_\pm$ and $\chi_\pm=\exp(F+K\pmb{\sigma}_1\cdot\hat{\mathbf{r}}\pmb{\sigma}_2\cdot\hat{\mathbf{r}})\eta_\pm$.
When considering only scalar and vector interactions, the resulting Schr\"odinger-like equation, after performing a Pauli reduction, takes the form:
\begin{eqnarray}
\label{Sch-like-DEq}
&&[\mathbf{p}^2 + \Phi_{\rm IS} + \pmb{\sigma}_1 \cdot \pmb{\sigma}_2 \Phi_{SS}+ \mathbf{L} \cdot(\pmb{\sigma}_1 + \pmb{\sigma}_2)\Phi_{SO} \nonumber\\
&&+ \pmb{\sigma}_1 \cdot \hat{\mathbf{r}} \pmb{\sigma}_2 \cdot \hat{\mathbf{r}}\mathbf{L} \cdot(\pmb{\sigma}_1 + \pmb{\sigma}_2)\Phi_{SOT} \nonumber\\
&&+ i \mathbf{L}\cdot \pmb{\sigma}_1 \times \pmb{\sigma}_2\Phi_{SOX}  + \mathbf{L} \cdot(\pmb{\sigma}_1 - \pmb{\sigma}_2)\Phi_{SOD} \nonumber\\
&&+ (3\pmb{\sigma}_1 \cdot \hat{\mathbf{r}} \pmb{\sigma}_2 \cdot \hat{\mathbf{r}} - \pmb{\sigma}_1 \cdot \pmb{\sigma}_2)\Phi_T] \psi_+ = b^2(w)\psi_+,
\end{eqnarray}
where $m_w = m_1m_2/w$, $\epsilon_w = (w^2 - m_1^2 - m_2^2)/2w$, and $b^2(w) = \epsilon_w^2 - m_w^2$. $\omega$ is the total energy. $\Phi_{\rm IS}=2m_w S + S^2 + 2 \epsilon_w A - A^2+\Phi_D$ is spin-independent potential with $\Phi_D$ the Dawin term. $S$ and $A$ are scalar (pseudoscalar) and time-like vector potentials, respectively. $\Phi_{SS}$, $\Phi_{SO}$, $\Phi_{SOT}$, $\Phi_{SOD}$, $\Phi_{SOX}$, and $\Phi_{T}$ are the spin–spin, spin–orbital and tensor interactions. The explicit forms of these terms can be found in Ref.~\cite{Crater:2008rt},
\begin{eqnarray}
\label{potential_terms}
\Phi_D &=& -{2F'(\cosh{2K} - 1)\over r} + F'^2 + K'^2 +m(r)\nonumber\\
&+&{2K'\sinh{2K} \over r} - \nabla^2 F - {2(\cosh{2K} - 1) \over r^2} ,\nonumber\\
\Phi_{SO} &=& -{F' \over r} - {F' (\cosh{2K} - 1) \over r} - {\cosh{2K} - 1 \over r^2} \nonumber\\
&+& {K' \sinh{2K} \over r},\nonumber\\
\Phi_{SOD} &=& l'(r)\cosh{2K} - q'(r)\sinh{2K},\nonumber\\ 
\Phi_{SOX} &=& q'(r)\cosh{2K} - l'(r)\sinh{2K},\nonumber\\
\Phi_{SS} &=& k(r) + {2K'\sinh{2K} \over 3r} - {2F'(\cosh{2K} - 1) \over 3r} \nonumber\\
&-& {2(\cosh{2K} - 1) \over 3r^2} + {2 F' K' \over 3} - {\nabla^2 K \over 3},\nonumber\\
\Phi_T &=& {1 \over 3} \Big[ n(r) + {3F'\cosh{2K} \over r} + {F' (\cosh{2K} - 1) \over r} \nonumber\\
&+& 2F'K' - {K'\sinh{2K} \over r} - {3K'(\cosh{2K} - 1) \over r}\nonumber\\
&-& \nabla^2 K + {3\sinh{2K} \over r^2} + {\cosh{2K} - 1 \over r^2}\Big],\nonumber\\
\Phi_{SOT} &=& -{K'(\cosh{2K} - 1) \over r} + {\sinh{2K} \over r^2} - {K' \over r} \nonumber\\
&+& {F'\sinh{2K} \over r},
\end{eqnarray}
where
\begin{eqnarray}
\label{abbreviation}
        \epsilon_1 &=& {\omega \over 2}+{m_1^2-m_2^2 \over 2\omega},\nonumber\\
        \epsilon_2 &=& {\omega \over 2}-{m_1^2-m_2^2 \over 2\omega},\nonumber\\
        M_1 &=& \sqrt{m_1^2+\exp(2\mathcal{G})(2m_{\omega}S+S^2)},\nonumber\\
        M_2&=& \sqrt{m_2^2+\exp(2\mathcal{G})(2m_{\omega}S+S^2)},\nonumber\\
        E_1 &=&\sqrt{\exp(2\mathcal{G})}(\epsilon_1-A), \nonumber\\
        E_2&=& \sqrt{\exp(2\mathcal{G})}(\epsilon_2-A),\nonumber\\
        F &=& {1 \over 2}\ln {E_2M_1 + E_1M_2 \over \epsilon_2 m_1 + \epsilon_1 m_2} - \mathcal{G},\nonumber\\
        \mathcal{G} &=& -{1 \over 2}\ln{( 1 - 2A/w )},\nonumber\\
        L &=& \ln{\left( M_1 + M_2 \over m_1 + m_2 \right)},\nonumber\\
        K &=& {\mathcal{G} + L \over 2}.   
\end{eqnarray}
The relevant derivatives are
\begin{eqnarray}
    L' &=& {w \over M_1M_2}\left( {S'(m_w + S) \over w - 2A} + {(2m_wS + S^2)A' \over (w - 2A)^2}\right),\nonumber\\
    \mathcal{G}' &=& {A' \over w - 2A},\nonumber\\
    F' &=& {(L' - \mathcal{G}')(E_2M_2 + E_1M_1) \over 2(E_2M_1 + E_1M_2)} - \mathcal{G}',\nonumber\\
    K' &=& {\mathcal{G}' + L' \over 2},
\end{eqnarray}
with $S'$ and $A'$ are the derivatives of the scalar and vector potentials, respectively. The Laplacian terms are 
\begin{eqnarray}
    \nabla^2 L &=&  {w \over M_1 M_2}\Big( 
    {4(2m_wS + S^2)A'^2 \over (w - 2A)^3}
   \nonumber\\ 
    &+& {4S'(m_w + S)A' + (2m_wS + S^2)\nabla^2A \over (w - 2A)^2} \nonumber\\
    &+&  {(m_w + S)\nabla^2 S + S'^2 \over w - 2A} \Big)-{L'^2(M_1^2 + M_2^2) \over M_1M_2},\nonumber\\
    \nabla^2 \mathcal{G} &=& {\nabla^2 A \over w - 2A} + 2\mathcal{G}'^2,\nonumber\\
    \nabla^2 F &=& {(\nabla^2 L - \nabla^2 \mathcal{G})(E_2M_2 + E_1M_1) \over 2(E_2M_1 + E_1M_2)} \nonumber\\
    &-& (L' - \mathcal{G}')^2 {(m_1^2 - m_2^2)^2 \over 2(E_2M_1 + E_1M_2)^2} - \nabla^2 \mathcal{G},\nonumber\\
    \nabla^2 K &=& {\nabla^2 \mathcal{G} + \nabla^2 L \over 2}.
\end{eqnarray}
Furthermore, other $r$-dependent functions have the form,
\begin{eqnarray}
        k(r) &=& {1 \over 3} \nabla^2(K + \mathcal{G}) - {2F'(\mathcal{G}' + K') \over 3} - {1\over 2}\mathcal{G}'^2,\nonumber\\
        n(r) &=& \nabla^2 K - {1 \over 2} \nabla^2 \mathcal{G} + {3(\mathcal{G}' - 2K') \over 2r} + F'(\mathcal{G}' - 2K'),\nonumber\\
        m(r) &=& -{1\over 2}\nabla^2 \mathcal{G} + {3\over 4}\mathcal{G}'^2 + \mathcal{G}'F' - K'^2,\nonumber\\
        l'(r) &=& -{1\over 2r}{E_2M_2 - E_1M_1 \over E_2M_1 + E_1M_2}(L' - \mathcal{G}'),\nonumber\\
        q'(r) &=& {1\over 2r}{E_1M_2 - E_2M_1 \over E_2M_1 + E_1M_2}(L' - \mathcal{G}').
\end{eqnarray}

From the above formulation, it is clear that the two-body Dirac equation~\eqref{Sch-like-DEq} depends on a single variable—the total energy $\omega$. The essential inputs to solve the equation are the masses of the two particles, $m_1$, $m_2$, along with the radial scalar potential $S(r)$, and time-like vector potential $A(r)$.
 
By separating the angular part in Eq.~\eqref{Sch-like-DEq}, 
the radial wave functions become governed by a set of coupled differential equations due to the presence of spin-dependent potentials. Specifically, the spin-singlet wave function $u_0$ and one of the spin-triplet $u_1^0$ with quantum numbers $^1J_j$ and $^3J_j$ are controlled by two coupled equations. Meanwhile, the other two triplet components, $u^+_1$ and $u^-_1$ associated quantum numbers $^3J_{j+1}$ and $^3J_{j-1}$ obey another set of coupled equations. In the special case where the two particles have equal masses, $m_1=m_2$, the tensor terms $\Phi_{SOD}$ and $\Phi_{SOX}$ vanish, causing the first set of coupled equations to decouple for the singlet states. For such singlet states (such as $^1S_0$, $^1P_1$, and $^1D_2$), the radial equation simplifies and can be written as follows:
\begin{eqnarray}
    \left[ -{\mathrm{d}^2 \over \mathrm{d}r^2} + {J(J+1) \over r^2} + \Phi \right]u_0(r) = b^2u_0(r),
\end{eqnarray}
where 
\begin{eqnarray}
    \Phi =  2m_wS + S^2 + 2 \epsilon_w A - A^2 + \Phi_D - 3\Phi_{SS}.
\label{eq.phis0}    
\end{eqnarray}
The $u_1^0$ wave satisfies (e.g. $^3P_1$ state),
\begin{eqnarray}
    \left[ -{\mathrm{d}^2 \over \mathrm{d}r^2} + {J(J+1) \over r^2} + \Phi \right]u_1^0(r) = b^2u_1^0(r),
\end{eqnarray}
where 
\begin{eqnarray}
    \Phi  &=&  2m_wS + S^2 + 2 \epsilon_w A - A^2 + \Phi_D +\Phi_{SS}\nonumber\\
    &-&2\Phi_{SO}+2\Phi_{T}-2\Phi_{SOT}.
\end{eqnarray}
The other two triplet states $u^+_1$ and $u^-_1$ remain coupled because of the non-zero tensor interactions $\Phi_T$ and $\Phi_{SOT}$ (e.g. $^3S_1$ and $^3D_1$ states are coupled with each other),
\begin{eqnarray}
    \left[ -{\mathrm{d}^2 \over \mathrm{d}r^2} + {J(J-1) \over r^2} + \Phi_{11} \right]u_1^++\Phi_{12}u_1^- &=& b^2u_1^+,\nonumber\\
    \left[ -{\mathrm{d}^2 \over \mathrm{d}r^2} + {(J+1)(J+2) \over r^2} + \Phi_{22} \right]u_1^-+\Phi_{21}u_1^+ &=& b^2u_1^-,\nonumber\\
\end{eqnarray}
where
\begin{eqnarray}
\Phi_{11} &=&  2m_wS + S^2 + 2 \epsilon_w A - A^2 + \Phi_D +\Phi_{SS}\nonumber\\
&+&2(J-1)\Phi_{SO}+{2(J-1)\over 2J+1}(\Phi_{SOT}-\Phi_{T}),\nonumber\\
\Phi_{12} &=& {2\sqrt{J(J+1)}\over 2J+1}(3\Phi_{T}-2(J+2)\Phi_{SOT}),\nonumber\\
\Phi_{22} &=&  2m_wS + S^2 + 2 \epsilon_w A - A^2 + \Phi_D +\Phi_{SS}\nonumber\\
&-&2(J+2)\Phi_{SO}+{2(J+2)\over 2J+1}(\Phi_{SOT}-\Phi_{T}),\nonumber\\
\Phi_{21} &=& {2\sqrt{J(J+1)}\over 2J+1}(3\Phi_{T}+2(J-1)\Phi_{SOT}).
\label{eq.phis1}
\end{eqnarray}

\subsection{Scattering phase shift}

The scattering problem is solved by integrating the radial Schr\"odinger equation. The scattering phase shift can be extracted by comparing the obtained scattering wavefunction with its known asymptotic form at large distances. Using the variable phase approach (VPA)~\cite{10.1119/1.1975005}, the phase shift $\delta_l$ for a given partial wave $l$ can be directly obtained from the radial potential through the differential equation:
\begin{eqnarray}
\delta'_l(r)=-k^{-1}V(r)[\cos\delta_l(r)j_l(kr)-\sin\delta_l(r)n_l(kr)]^2,
\label{eq.vpa}
\end{eqnarray}
where $j_l$ and $n_l$ are the spherical Bessel and Neumann functions, respectively. $j_0(x)=\sin x$, $j_1(x)=-\cos x+(\sin x)/x$. $n_0(x)=-\cos x$, $n_1(x)=-\sin x-(\cos x)/x$. $k$ is the relative momentum.
This equation is also known as the phase equation. It is a first-order nonlinear differential equation whose solution directly yields the scattering phase shift in the asymptotic limit. Specifically, the phase shift for the $l$-th partial wave is given by the limit
\begin{eqnarray}
\delta_l=\lim_{r\to \infty}\delta_l(r).
\end{eqnarray}
where $\delta_l(r)$ is the solution of the phase equation at radius $r$.

In low energy scattering, the $S$-wave scattering is dominate. We can approximately let $l = 0$, which lead to $S$-wave variable phase equation, reads
\begin{eqnarray}\label{eq.vpa-sWave}
    \delta'_0(r) = -k^{-1} 2\mu V(r) \sin^2[kr + \delta_0(r)],
    \label{lab-phase}
\end{eqnarray}
where $\mu=m_1m_2/(m_1+m_2)=m/2$ is the reduced mass of two particles. The boundary condition can be taken as $\delta_0(r=0)=0$.

In the relativistic case, we got the similar equation for the uncoupled cases,
\begin{eqnarray}\label{eq.vpa-sWave}
\delta'_0(r) = -b^{-1} \Phi(r,\omega) \sin^2[br + \delta_0(r)].
\end{eqnarray}
The main difference is that the relativistic potential depends explicitly on the energy or equivalently on the input value of $b$. Therefore, for a given $b$, we must solve self-consistently for the total energy $\omega$ and the energies of each particle,
\begin{eqnarray}
w = 2\sqrt{b^2 +m^2},\quad \epsilon_1 = \epsilon_2 = {w \over 2},
\end{eqnarray}
which also enter into the potential.

For the coupled case, we need to solve the coupled phase equation to obtain the phase shifts~\cite{Liu:2002cn}, 
\begin{eqnarray}
\delta'_S(r)&=& -b^{-1} \Big[(\Phi_{11}+{J(J-1)\over r^2})\cos^2 \varepsilon(r) \nonumber\\
&+&(\Phi_{22}(r)+{(J+1)(J+2)\over r^2})\sin^2\varepsilon(r) \nonumber\\
&+ &\Phi_{12}\sin 2\varepsilon(r) \Big]\times\sin^2[br + \delta_S(r)], \nonumber\\
\delta'_D(r)&=& -b^{-1} \Big[(\Phi_{22}+{(J+1)(J+2)\over r^2})\cos^2 \varepsilon(r) \nonumber\\
&+&(\Phi_{11}(r)+{J(J-1)\over r^2})\sin^2\varepsilon(r) \nonumber\\
&+& \Phi_{21}\sin 2\varepsilon(r) \Big]\times \sin^2[br + \delta_D(r)], \nonumber\\
\varepsilon'(r)&=& (b\sin[\delta_S(r)-\delta_D(r)])^{-1} \Big[{1\over 2}(\Phi_{11}+{J(J-1)\over r^2} \nonumber\\
&-&\Phi_{22}-{(J+1)(J+2)\over r^2})\sin2\varepsilon(r)-\Phi_{12}\cos2\varepsilon(r) \Big ]\nonumber\\
&\times& \sin[br + \delta_S(r)]\sin[br + \delta_D(r)].
\label{eq.phtriplet}
\end{eqnarray}
These coupled ordinary differential equations can be solved numerically using the fourth-order Runge-Kutta method. The solution provides the scattering phase shifts 
for the $S$ and $D-$wave channels, denoted as, $\delta_S$ and $\delta_D$.

\subsection{Correlation function}

As mentioned in the introduction, the correlation function admits an analytical expression under the assumption of a Gaussian source function and short-range potential approximation. The Gaussian form of the emission source is given by
\begin{eqnarray}
 S(r)={1\over (4\pi R^2)^{3/2}}\exp \left (-{r^2\over 4R^2}\right )
\end{eqnarray}
with a constant parameter $R$ which described the size of the emission source. For proton-proton collisions, $R\approx1.0~\rm fm$~\cite{Wang:2024bpl,ALICE:2020ibs} and for heavy-ion collisions, $\sim 5~\rm fm$ or even larger. In this study, we take $R=1.0~\rm fm$. The asymptotic form of the scattering wavefunction can be expressed as,
\begin{eqnarray}
\psi_{\rm asy}(r,k)= N \left[{\sin kr \over kr}+f(k){e^{ikr}\over r} \right],
\end{eqnarray}
where $f(k)$ is scattering phase shift and for a short range potential and low energy scatterings, the phase shift has a form,
\begin{eqnarray}
f(k)=\left(-{1\over a_0}+{1\over 2}r_{\rm eff}k^2-ik \right)^{-1}.
\end{eqnarray}
The scattering length $a_0$ and effective range $r_{\rm eff}$ can be obtained by expanding the scattering phase shift,
\begin{eqnarray}
k\cot \delta(k)=-{1\over a_0}+{1\over 2}r_{\rm eff}k^2+\mathcal{O}(k^4).
\label{eq.aandr}
\end{eqnarray}
For the spin singlet state, the scattering phase shift $\delta$ can be obtained by solving the Eq.~\eqref{eq.vpa-sWave}.
For the spin triplet state, by solving the coupled phase equation, Eq.~\eqref{eq.phtriplet}, we obtained the phase shift for both $S$ and $D-$wave, named $\delta_S$ and $\delta_D$, respectively. These phase shifts can be expanded at low energies to extract the scattering length $a_0$ and effective range $r_{\rm eff}$ for both $S$ and $D-$wave. However, in low-energy scattering processes, the dominant contribution comes from the $S-$wave phase shift $\delta_S$, while the influence of the $D-$wave component is negligible.

\begin{figure}[!htb]
    \centering
    \includegraphics[width=0.45\textwidth]{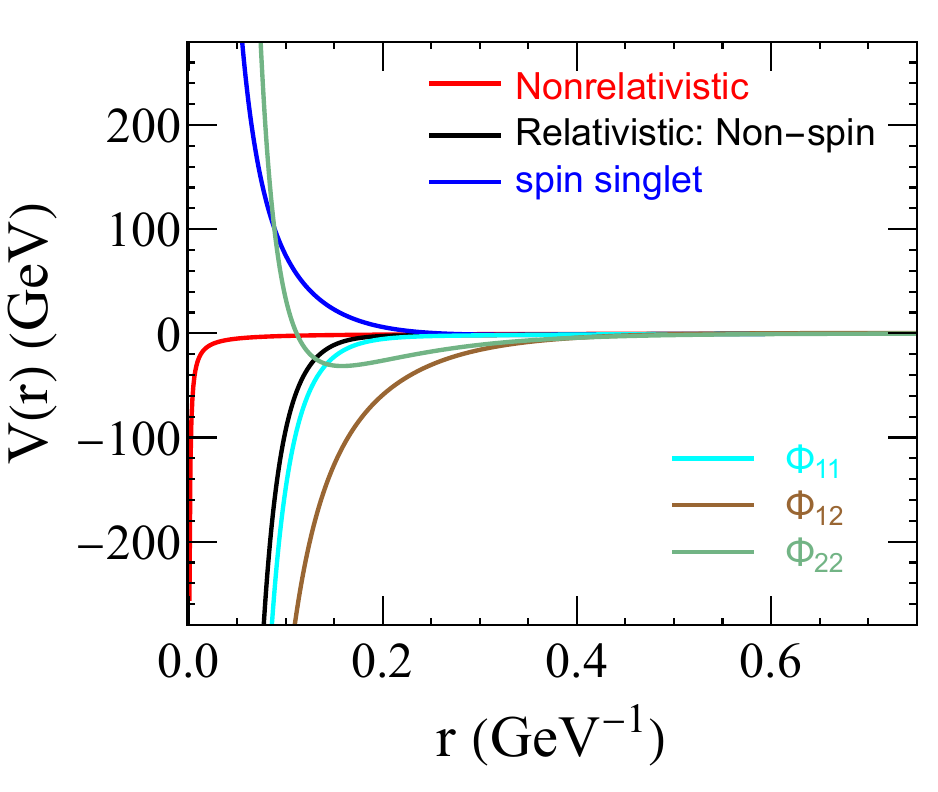}
    \caption{The non-relativistic potential (red line) and relativistic potentials without spin-dependent interactions (black dashed line). For the relativistic potential, we take $\omega=2~\rm GeV$ and $m=0.938~\rm GeV$, which corresponds to $b=0~\rm GeV$.}
    \label{fig.potential_Yukawa}
\end{figure}

The correlation function can be expressed as~\cite{Lednicky:1981su,Ohnishi:2016elb},
\begin{eqnarray}
C(k)&=&1+{|f(k)|^2\over 2R^2}F_3\left({r_{\rm eff}\over R}\right)+{2{\rm Re} f(k)\over\sqrt{\pi}R}F_1(2x)\nonumber\\
&-&{{\rm Im} f(k)\over R}F_2(2x),
\label{eq.corrf}
\end{eqnarray}
where $x=kR$. $F_1=(1+c_1x^2 +c_2x^4 +c_3x^6)/(1+(c_1 +2/3)x^2 +c_4x^4 +c_5x^6 +c_3x^8)$ with $(c_1,...,c_5)=(0.123,0.0376,0.0107,0.304,0.0617)$. $F_2(x)=(1-e^{-x^2})/x$ and $F_3(x)=1-x/(2\sqrt{\pi})$.

\section{Results}
\label{sec.res}

In this section, we take proton-proton scattering as an example to investigate the relativistic corrections to the phase shift and the correlation function, both of which are experimentally accessible observables. The proton-proton interaction potential has been constructed by fitting nucleon-nucleon scattering data. Several phenomenological potentials have been developed for this purpose, among which the Reid potential~\cite{Day:1981zz,Stoks:1994wp} and the \textit{Argonne}-18 potential~\cite{Wiringa:1994wb} are widely used.
\begin{figure}[!htb]
    \centering
    \includegraphics[width=0.45\textwidth]{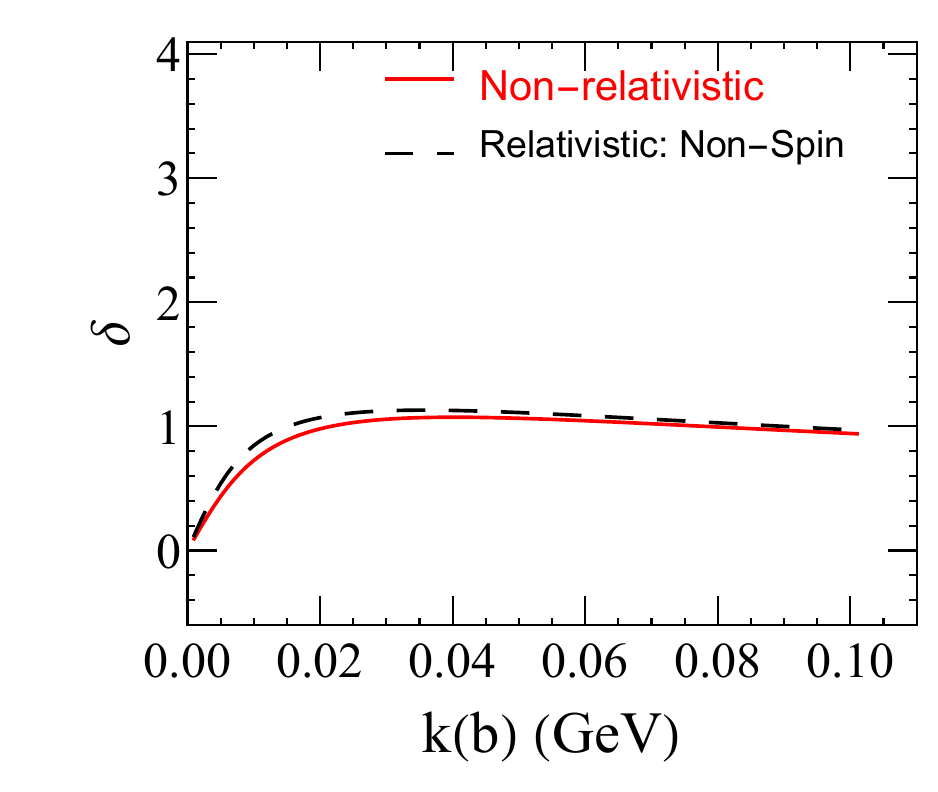}\\
    \includegraphics[width=0.45\textwidth]{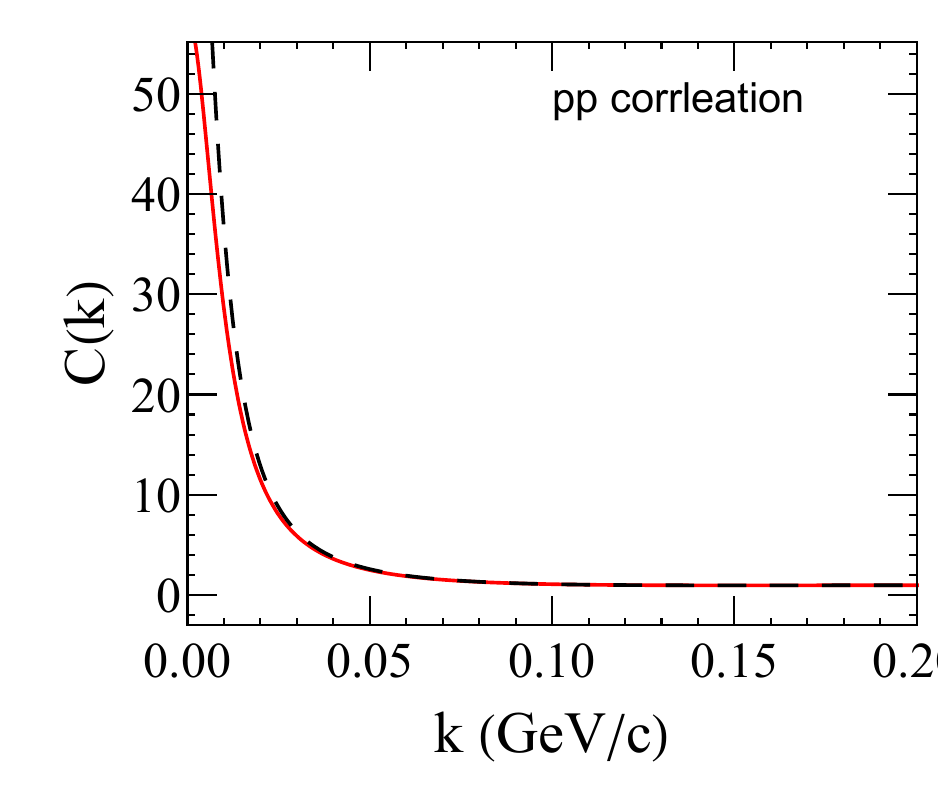}
    \caption{(Top panel) Scattering phase as a function of $k$. (Bottom panel) The correlation function. The black dashed lines are the non-relativistic case, while the red solid lines correspond to the relativistic case without spin-dependent potentials.}
    \label{fig.nonrela}
\end{figure}
\begin{figure}[!htb]
    \centering
    \includegraphics[width=0.45\textwidth]{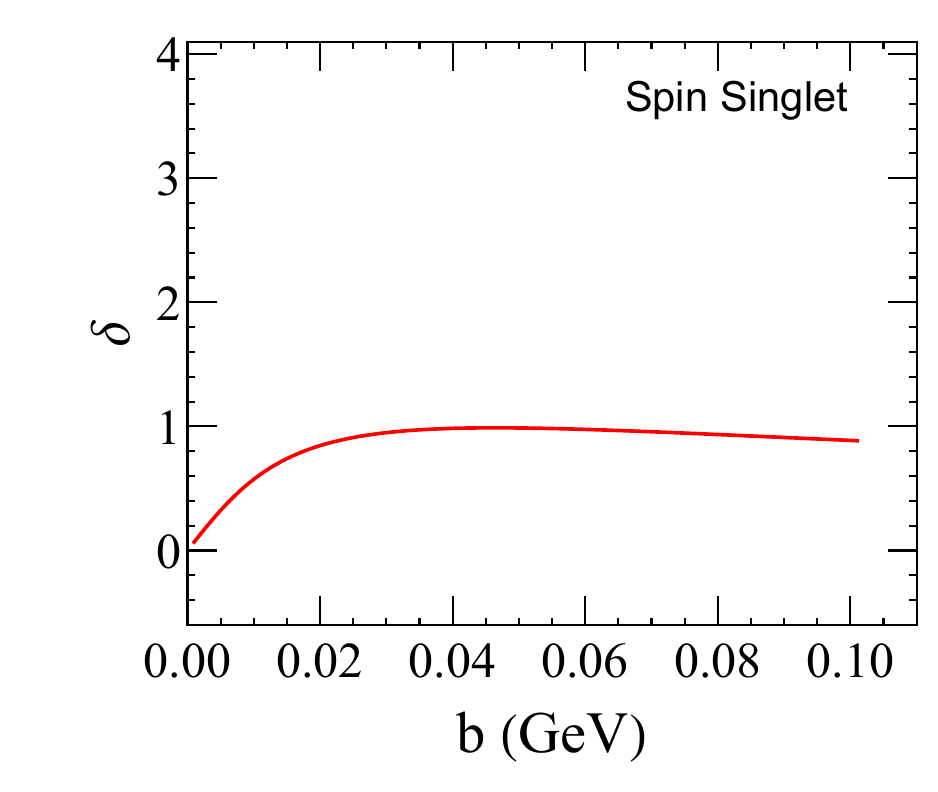}
    \includegraphics[width=0.45\textwidth]{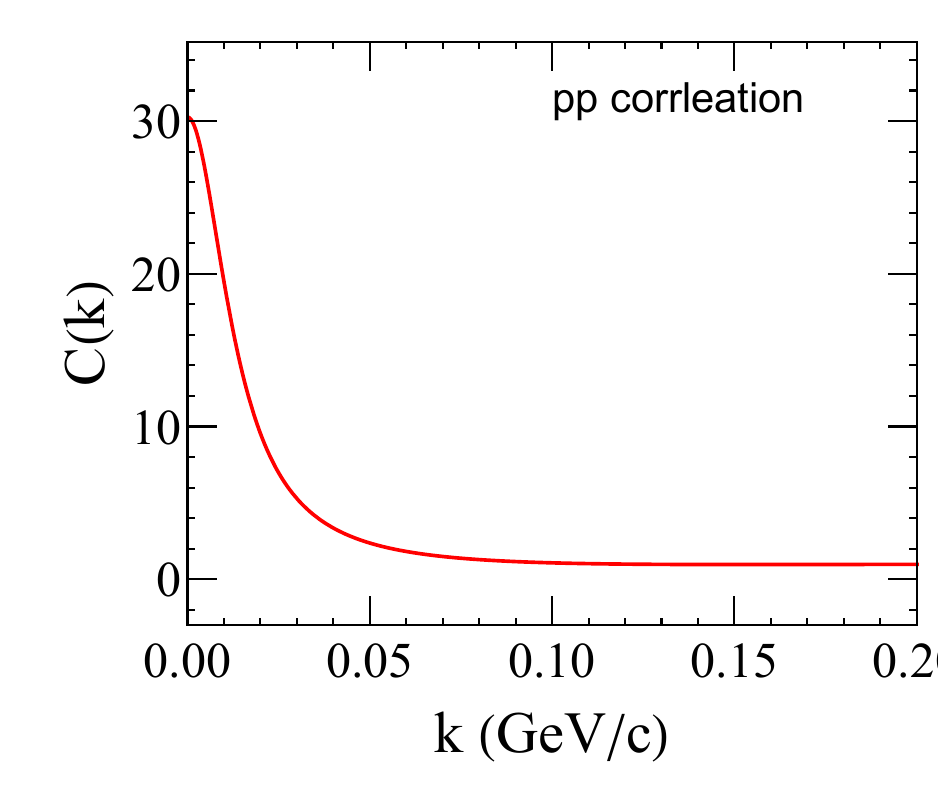}
    \caption{Spin-singlet state: (top panel) Scattering phase as a function of $k$. (Bottom panel) The correlation function.} 
    \label{fig.s0}
\end{figure}
As a simple estimation, we adopt a Yukawa-type potential, which assumes that the interaction between two protons is mediated by the exchange of a $\pi$-meson. For simplicity, the Coulomb repulsion between the protons and quantum statistics are neglected as a first approximation. The potential takes the following form:
\begin{align}
\label{NR_potential}
    V_{\rm pp}(r) = -V_0{e^{-m_\pi r}\over m_\pi r},
\end{align}
where the depth $V_0$ of the potential is related to the dimensionless pseudovector pion-nucleon coupling constant $f_\pi$, $V_0=m_\pi f_\pi^2$. For two protons, they interact via the exchange of a neutral pion. The two parameters $f_\pi$ and $m_\pi$ can be fixed by comparing to the nuclear scattering length $a{\rm pp}$ and effective range $r_{\rm pp}$ for proton-proton scattering. We choose the same value as obtained in Ref.~\cite{Babenko:2016idp}, which gives $m_\pi=0.165~\rm GeV$ and $f_\pi= 0.2707$. 

The Yukawa-type proton-proton (p-p) potential is implemented as a scalar or pseudoscalar interaction $S$, while the time-like vector interaction $A$ is set to zero in this case. To incorporate other forms of interactions, such as vector and tensor interactions, one would need to extend the Dirac equation accordingly, as demonstrated in Ref.~\cite{10.1063/1.532311}.

The non-relativistic form of the p-p potential is shown as the red line in Fig.~\ref{fig.potential_Yukawa}. In the relativistic case, the potential receives additional contributions from both kinetic and spin-dependent terms, as shown in Eq.~\eqref{Sch-like-DEq}. The relativistic potential without spin-dependent terms—meaning only including the Darwin term $\Phi_D$, is shown as the black line in Fig.~\ref{fig.potential_Yukawa}. We observe that the relativistic correction leads to a deeper potential well compared to the non-relativistic case.
The magnitude and form of the spin-dependent potential depend on the spin configuration of the two particles. For the spin-singlet state, only the spin-spin interaction $\Phi_{SS}$ contributes, as given in Eq.~\eqref{eq.phis0}. For this case, where $\bm \sigma_1\cdot \bm \sigma_2=-3$, the total singlet potential becomes strong repulsive, as illustrated by the blue line in Fig.~\ref{fig.potential_Yukawa}.
For the spin-triplet state, the two-body Dirac equation reduces to a set of coupled equations, and the relevant potentials $\Phi_{11}$, $\Phi_{12}$, and $\Phi_{22}$ appear, as defined in Eq.~\eqref{eq.phis1} and shown in Fig.~\ref{fig.potential_Yukawa} with different colors.
Additionally, we have verified that the relativistic results smoothly reduce to the non-relativistic limit as the particle masses $m\to \infty$. 

\begin{figure}[!htb]
    \centering
    \includegraphics[width=0.45\textwidth]{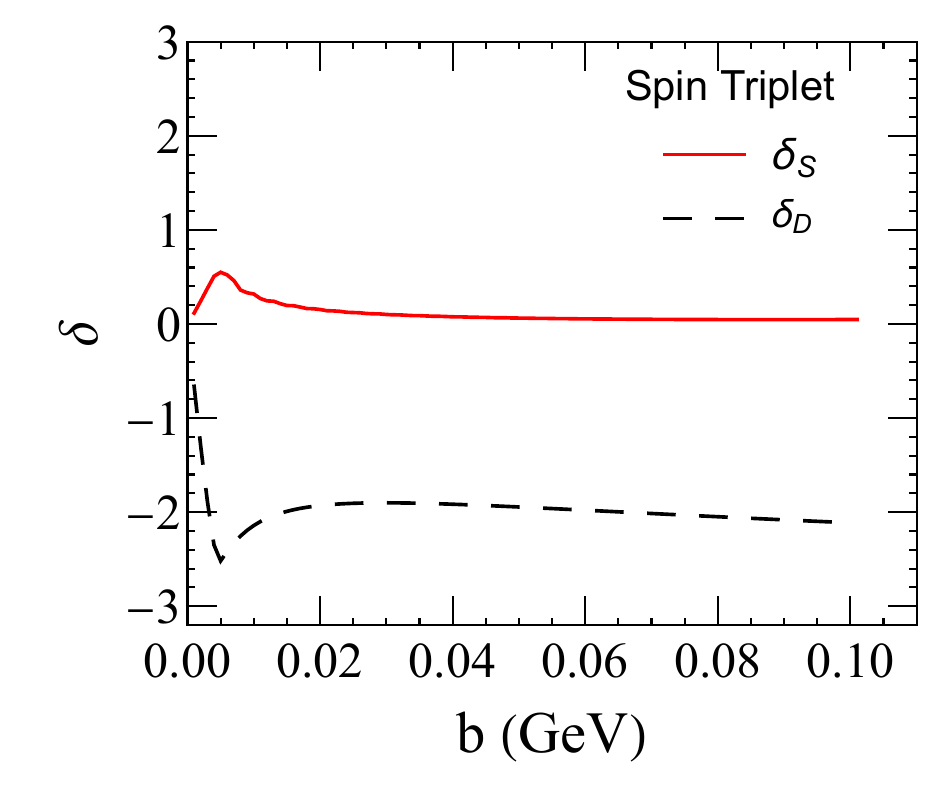}
    \includegraphics[width=0.45\textwidth]{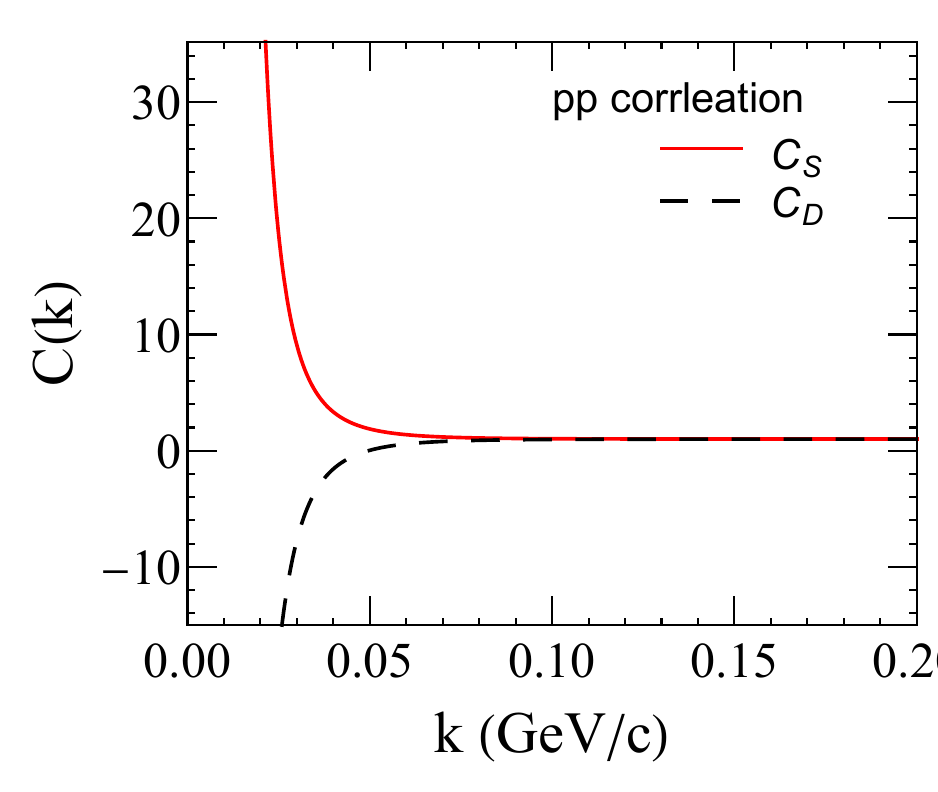}
    \caption{Spin-triplet state: (top panel) scattering phase as a function of $b$. The red line and black-dashed lines correspond to the $S-$wave ($\delta_S$) and $D-$wave ($\delta_D$) components, respectively. (Bottom panel) The correlation functions of $S-$wave ($C_S$) and $D-$wave ($C_D$)  components.}
    \label{fig.s1}
\end{figure}

We now turn to the calculation of the phase shifts for both the non-relativistic scenario and the relativistic cases, including: (i) the relativistic case without spin-dependent potentials, (ii) the spin-singlet state, and (iii) the spin-triplet state. In this study, we focus on low-energy scattering, where the dominant contribution arises from the $S-$wave component. By taking the large-distance limit of the phase function, we extract the scattering phase for a given momentum as $\lim_{r \to \infty}\delta_S(r)=\delta_S$. Repeating this procedure for various momenta $k$, we obtain the scattering phase shift $\delta_S(k)$. 
The numerical results are presented in Fig.~\ref{fig.nonrela} for non-relativistic and relativistic w/o spin potential. Fig.~\ref{fig.s0} for spin-singlet state. Fig.~\ref{fig.s1} for spin-triplet state with $S-$wave and $D-$wave components, $\delta_S$ and $\delta_D$. 
The scattering phase $k\cot \delta(k)$ can be expanded around $k=0$ to obtain the scattering length and effective range,
as shown in Eq.~\eqref{eq.aandr}. The values are shown in Table~\ref{table1}.

First, we observe only a slight difference between the scattering phases of the non-relativistic and relativistic cases without spin-dependent potentials, as shown in Fig.~\ref{fig.nonrela}. The extracted scattering length and effective range for the non-relativistic case are consistent with previous results~\cite{Botermans:1990qi}. In the relativistic case, the Darwin term introduces an additional short-range attraction in the central potential. Consequently, the scattering length becomes more positive or more negative, depending on the specific form of the potential, but generally increases in magnitude. Meanwhile, the effective range is slightly reduced, as shown clearly in Table~\ref{table1}.

For the spin-singlet state, the inclusion of the spin-spin interaction significantly reduces the depth of the potential well (a very shallow potential well). As a result, the scattering length decreases, and the effective range increases compared to the spin-independent case. In the spin-triplet state, the $S$ and $D-$waves are coupled through tensor interactions. At low energies, this coupling is relatively weak. However, the presence of a short-range repulsive core in the $\Phi_{22}$ potential distorts the wave function, which manifests in the phase shift as a negative effective range.

\begin{figure}[!htb]
    \centering
    \includegraphics[width=0.45\textwidth]{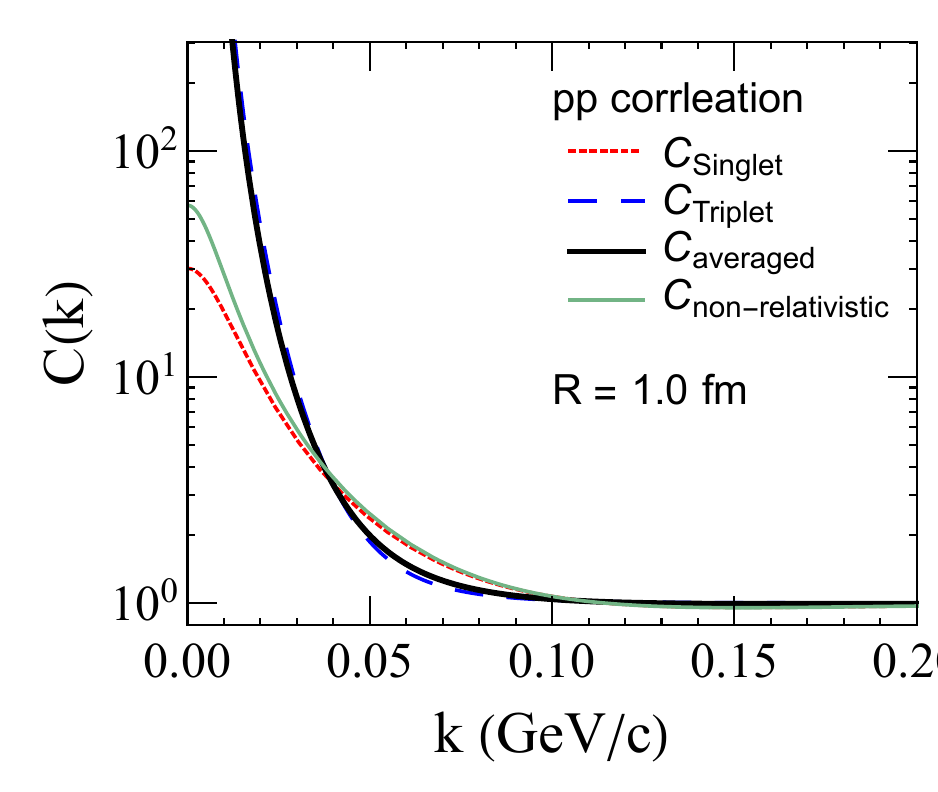}  
    \caption{The correlation function of spin-singlet (red-dotted line), triplet (blue-dashed line), spin-averaged case (thick-black line), and the non-relativistic case (thin-green line).}
    \label{fig.total}
\end{figure}

\begin{table}
	\renewcommand\arraystretch{1.5}
	\setlength{\tabcolsep}{5.mm}
	\begin{tabular}{c|c|c}
		\toprule[1pt]\toprule[1pt]
        \multicolumn{1}{c|}{}&
	\multicolumn{1}{c|}{$a_0$ \rm (fm) }& \multicolumn{1}{c}{$r_{\rm eff}$ \rm (fm)} 
        \tabularnewline
		\midrule[1pt]
	\rm non-relativistic & -18.69 & 2.82  
        \tabularnewline
        \rm relativistic w/o spin & -24.15 & 2.71 
        \tabularnewline
	\rm spin singlet & -13.43 & 2.99
        \tabularnewline
        \rm spin triplet & -22.10 &  -129.35
        \tabularnewline
		\bottomrule[1pt]
	\end{tabular}
	\caption{The $S-$wave scattering length $a_0$ and effective range $r_{\rm eff}$ in different cases.}
	\label{table1}
\end{table}
Next, we investigate the two-particle correlation functions based on the previously discussed LL model. The momentum-dependent correlation functions are shown in Fig.~\ref{fig.nonrela} for the non-relativistic case and the relativistic scenario without spin-dependent potentials, in Fig.~\ref{fig.s0} for the spin-singlet state, and in Fig.~\ref{fig.s1} for the spin-triplet state.
From Fig.~\ref{fig.nonrela}, we observe that the correlation function is slightly enhanced in the relativistic case compared to the non-relativistic one. This enhancement originates from the deeper effective potential well introduced by relativistic corrections, which increases the overlap between the interacting pair and the emission source.
In the spin-singlet state, shown in Fig.~\ref{fig.s0}, the presence of the spin-spin interaction reduces the depth of the potential, leading to weaker correlations.
For the spin-triplet states, illustrated in Fig.~\ref{fig.s1}, there are two distinct contributions to the correlation function, corresponding to the $S$ and $D-$wave components, $C_S$ and $C_D$, depicted by the red-solid and black-dashed lines, respectively. The $S-$wave contribution exhibits a positive correlation at low relative momentum, consistent with the attractive nature of the $S$-wave interaction. In contrast, the $D-$wave contribution shows a negative correlation at small $k$, which can be attributed to the repulsive effect of the centrifugal barrier, proportional to $(J+1)(J+2)/r^2$.

In experimental measurements, distinguishing the spin state of the two-particle system is generally not feasible. As a result, the observed correlation function corresponds to a spin-averaged quantity. Assuming the absence of polarization, the contributions from different spin states are weighted according to their statistical spin factors. Specifically, the spin-averaged correlation function is expressed as,
\begin{eqnarray}
C_{\rm averaged}(k)={1\over 4}C_{\rm Singlet}(k)+{3\over 4}C_{\rm Triplet}(k),
\end{eqnarray}
where $C_{\rm Singlet}$ are $C_{\rm Triplet}$ the correlation functions for the spin-singlet and spin-triplet states, respectively. The spin-averaged correlation function is shown in Fig.~\ref{fig.total} with thick-black line. Compared to the non-relativistic scenario $C_{\rm non-relativistic}$, it is evident that the inclusion of relativistic corrections enhances the two-nucleon correlation function due to the stronger effective interaction. This highlights the importance of considering relativistic effects in the study of hadron-hadron correlations. Conversely, when extracting information about the hadron emission source and the underlying hadronic interaction from experimental correlation data, it is essential to account for these relativistic corrections.

\section{Summary}
\label{sec.sum}
In summary, we have solved the two-body scattering problem within the relativistic two-body Dirac equation framework. The scattering phase shifts were obtained by numerically solving the variable phase equation. From the phase shifts, we extracted key scattering parameters such as the scattering length and effective interaction range. Utilizing these parameters, we computed the two-particle correlation functions and systematically compared results from the non-relativistic case, the relativistic scenario without spin-dependent potentials, the relativistic spin-singlet state, and the relativistic spin-triplet state.

Our analysis shows that relativistic effects play a significant role, notably enhancing the proton-proton correlation function. Although this study focused on proton-proton scattering, we anticipate that relativistic corrections will be even more pronounced in lighter hadronic systems, such as $K-K$, $p-K$ pairs. Future work will extend this framework to incorporate vector potentials, which arise naturally from $\omega$ and $\rho$ meson exchanges, thereby providing a more comprehensive description of the hadronic interaction.

\vspace{1cm}
{\bf Acknowledgement:} BC is supported by the National Natural Science Foundation of China (NSFC) under Grant No.
12175165. JX is supported by the Helmholtz Research Academy Hessen for FAIR (HFHF).

\bibliography{Ref}

\end{document}